\documentclass[showpacs,twocolumn,preprintnumbers,amsmath,amssymb,prl,epsf,superscriptaddress]{revtex4-1}
\usepackage{graphicx}
\usepackage{nicefrac}
\usepackage{color}
\usepackage{lineno}

\begin{document}
\title{Indefinite-mean Pareto photon distribution from amplified quantum noise}
\author{Mathieu Manceau}
\thanks{K.Yu.S. and M.M. contributed equally.}
\affiliation{Max Planck Institute for the Science of Light, Staudtstra\ss{}e 2, 91058 Erlangen, Germany}
\affiliation{Universit\'e Paris 13, Sorbonne Paris Cit\'e, Laboratoire de Physique des Lasers, 93430 Villetaneuse, France}
\author{Kirill Yu.~Spasibko}
\email{kirill.spasibko@mpl.mpg.de}
\affiliation{Max Planck Institute for the Science of Light, Staudtstra\ss{}e 2, 91058 Erlangen, Germany}
\affiliation{University of Erlangen-N\"urnberg, Staudtstra\ss{}e 7/B2, 91058 Erlangen, Germany}
\author{Gerd Leuchs}
\affiliation{Max Planck Institute for the Science of Light, Staudtstra\ss{}e 2, 91058 Erlangen, Germany}
\affiliation{University of Erlangen-N\"urnberg, Staudtstra\ss{}e 7/B2, 91058 Erlangen, Germany}
\author{Radim Filip}
\affiliation{Department of Optics, Palacky University, 77146 Olomouc, Czech Republic}
\author{Maria~V.~Chekhova}
\affiliation{Max Planck Institute for the Science of Light, Staudtstra\ss{}e 2, 91058 Erlangen, Germany} 
\affiliation{University of Erlangen-N\"urnberg, Staudtstra\ss{}e 7/B2, 91058 Erlangen, Germany}
\affiliation{Department of Physics, M.V.Lomonosov Moscow State University, \\ Leninskie Gory,
119991 Moscow, Russia}

\begin{abstract}
Extreme events appear in many physics phenomena, whenever the probability distribution has a `heavy tail', differing very much from the equilibrium one. Most unusual are the cases of power-law (Pareto) probability distributions. Among their many manifestations in physics, from `rogue waves' in the ocean to L\'{e}vy flights in random walks, Pareto dependences can follow very different power laws. For some outstanding cases the power exponents are less than 2, leading to indefinite mean values, let alone higher moments. Here we present the first evidence of indefinite-mean Pareto distribution of photon numbers at the output of nonlinear effects pumped by parametrically amplified vacuum noise, known as bright squeezed vacuum (BSV). We observe a Pareto distribution with power exponent 1.31 when BSV is used as a pump for supercontinuum generation, and other heavy-tailed distributions (however with definite moments) when it pumps optical harmonics generation. Unlike in other fields, we can flexibly control the Pareto exponent by changing the experimental parameters. This extremely fluctuating light is interesting for ghost imaging and quantum thermodynamics as a resource to produce more efficiently non-equilibrium states by single-photon subtraction, the latter we demonstrate experimentally.
\end{abstract}

\maketitle \narrowtext

`Common sense' or, rather, the central limit theorem tells us that the probability distributions of random physical values tend to be Gaussian. The more surprising are deviations from this tendency, from the behaviour of wave height in the ocean to the statistics of solar flares. Examples are `rogue waves' in the ocean~\cite{Kharif2003} and their analogues in 
nonlinear optics~\cite{Solli2007,Akhmediev2016}, L\'{e}vy flights~\cite{Barthelemy2008,Mercadier2009}, and other numerous examples in astrophysics~\cite{Lu1991}, geophysics~\cite{Taubert2018}, condensed matter physics~\cite{Stefani2009,Campi2015}, etc~\cite{Clauset2009}. Following oceanology, the term `rogue waves' denotes events whose magnitude considerably exceeds the ones expected from Gaussian statistics~\cite{Akhmediev2010}. A distribution with a high probability of such events is said to have a `heavy tail'~\cite{Foss2013}. 

An example of extreme heavy-tailed distribution is the Pareto one, typically describing the statistics of income and wealth~\cite{Yakovenko2009} and scaling as the power law with an exponent $1+k$: $P(x)\propto x^{-(1+k)}$. For any $k$, certain statistical moments of this distribution do not converge. At $k<1$, even the mean value diverges. A power-law 
distribution can appear due to the exponential amplification of an initially broad distribution~\cite{Newman2005}. 
As we show below, a similar phenomenon occurs 
when the exponential rate of frequency conversion is governed by the amplified vacuum noise.  

Vacuum noise, also called zero-point vacuum fluctuations, originates from the non-commutativity of photon creation and annihilation 
operators~\cite{Klyshko1988}. It causes spontaneous transitions in atoms as well as spontaneous 
parametric down-conversion (PDC) and four-wave mixing (FWM). In other words, PDC and FWM `visualize' the vacuum noise by parametrically 
amplifying it. Because parametric amplification is accompanied by squeezing, the quantum state of light produced through PDC and FWM is 
called squeezed vacuum. This state manifests quadrature and photon-number squeezing~\cite{Jedrkiewic2004,Iskhakov2009,Vahlbruch2016}, but in the context of this work, its most important feature is superbunching~\cite{Boitier2011,Iskhakov2012}, i.e., photon-number fluctuations stronger than the ones of thermal light~\footnote{Superbunching of BSV originates from the fact that its Wigner function is stretched, one of the quadratures being strongly anti-squeezed and the other one, squeezed. It is this antisqueezing that is responsible for enhanced intensity fluctuations.}.
\begin{figure}[h]
\begin{center}
\includegraphics[width=0.5\textwidth]{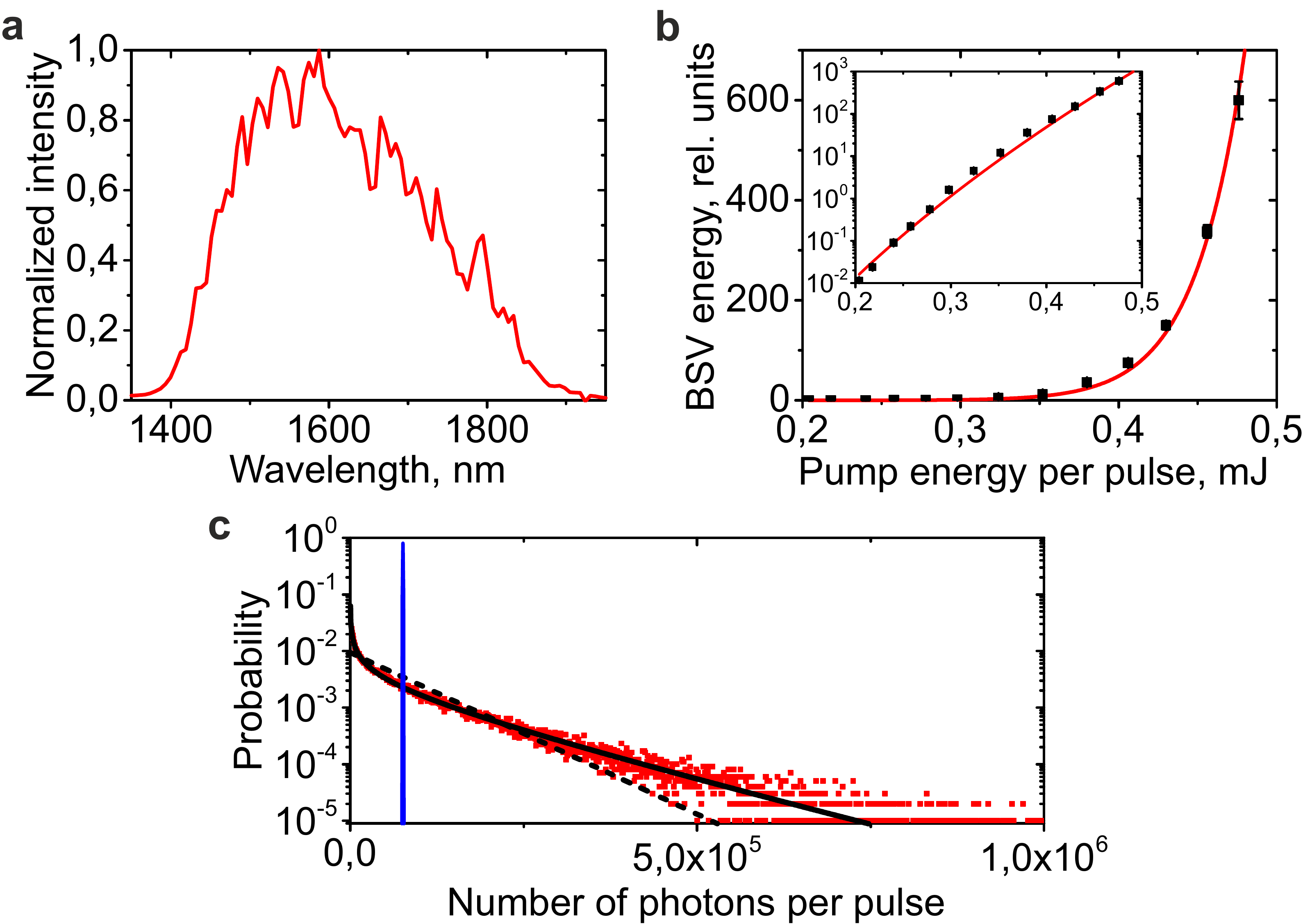}
\caption{Bright squeezed vacuum we use in experiment: a, the spectrum under pumping at $800$ nm. b, exponential dependence of the output energy per pulse on the input one in the linear and log-linear scales (inset). The red line is the fit according to the corresponding hyperbolic 
dependence~\cite{Iskhakov2009}. c, the probability distribution of the photon number per pulse, experimental (red points) and theoretical (Eq.~\eqref{eq:P_sb}, solid black line) compared to the thermal (dashed black line) and Poissonian (blue line) distributions with the same mean value.}\label{BSV} 
\end{center}
\end{figure}

Through PDC from strong picosecond pulses (see the Supplementary Information and Ref.~\cite{Spasibko2017}), we produce squeezed vacuum that is bright enough to pump nonlinear effects such as the generation of optical harmonics or supercontinuum. The spectral width of the resulting BSV exceeds 35~THz for pumping at $800$ nm (Fig.~\ref{BSV}a). Within this band, the vacuum noise is exponentially amplified to a brightness of $N=\sinh^2(G)$ photons per mode, where the parametric gain $G$ scales with the pump laser field amplitude. Figure~\ref{BSV}b shows this exponential amplification, with $G$ reaching $15.3\pm0.5$. As a result, even after a narrowband filtering the mean photon number per pulse is still large. Around this mean, the photon number has a very broad distribution (Fig.~\ref{BSV}c), whose envelope is given by~\cite{Akhmanov1981, Leuchs2015}
\begin{equation}
P_{B}(N_B) = \frac{1}{\sqrt{2\pi\langle N_B\rangle N_B}}e^{-\frac{N_B}{2\langle N_B\rangle}}.
\label{eq:P_sb}
\end{equation}
It is broader than not only a Poissonian distribution (blue solid line), typical for a shot noise limited laser, but also a negative-exponential one (dashed black line), typical for light with thermal statistics. 

\begin{figure}[h]
\begin{center}
\includegraphics[width=0.4\textwidth]{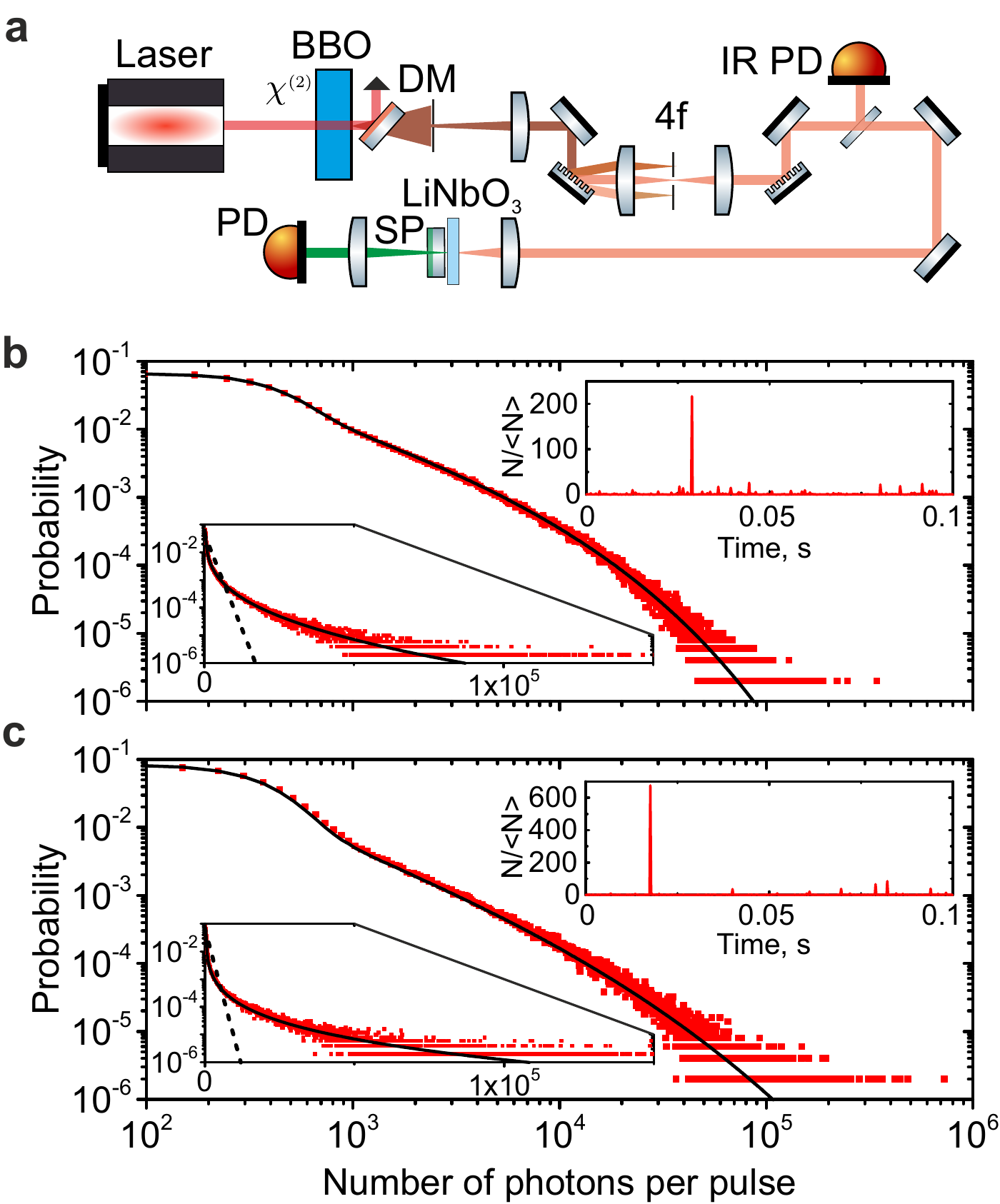}
\caption{Optical harmonics generated from squeezed vacuum. a, the experimental setup: after the BBO crystal, the pump is cut off by a 
dichroic mirror (DM) and BSV is filtered spatially by a slit and spectrally by a 4f system; its statistics are analyzed by an infrared photodetector (IR PD); optical harmonics are generated in a lithium niobate crystal (LiNbO$_3$), filtered with a short-pass filter (SP) and measured using a visible photodetector (PD).  b,c, experimental (red points) and theoretical (black lines) probability distributions of photon number per pulse for SH (b) and TH (c). Bottom-left insets: the same distributions in log-linear scale highlight strong deviations from the thermal light distribution (dashed line) with the same mean. Top-right insets: typical time traces showing pulses with `extreme' heights, normalized to the mean value.} \label{harmonics}
\end{center}
\end{figure}

We further use BSV to pump optical harmonic generation (Fig.~\ref{harmonics}a).  Panels b,c display the probability distributions for the pulses of second harmonic (SH) and third harmonic (TH). The log-log scale stresses that within a certain range, the scaling is close to a power law. However, at large photon numbers the decay becomes faster and, as a result, all moments of this distribution exist. The bottom-left insets show the same distribution in log-linear scale, demonstrating a strong deviation from the thermal-light distribution (dashed line). 

This behavior can be explained by the fact that the number of photons $N_{n\omega}$ in the $n$th harmonic is a power function of the number $N_\omega$ of photons in the fundamental radiation,
\begin{equation}
N_{n\omega} = KN_\omega^n,
\label{N_nw}
\end{equation}
where $K$ depends on the conversion efficiency. 
The probability distribution $P_{n\omega} (N_{n\omega})$ for the harmonic radiation can be obtained from the one for the 
fundamental radiation $P_{\omega}(N_\omega)$
as $P_{n\omega}(N_{n\omega})dN_{n\omega}=P_{\omega}(N_\omega)dN_\omega$:
\begin{equation}
P_{n\omega}(N_{n\omega}) = \frac{P_{\omega}\left(\sqrt[n]{N_{n\omega}/K}\right)}{n\sqrt[n]{K}N_{n\omega}^{1-\nicefrac{1}{n}}}.
\label{P_nw}
\end{equation}
Taking the photon-number distribution for BSV in the form (\ref{eq:P_sb}), for its $n$th harmonic we get
\begin{eqnarray}
&&P_{n\omega}(N_{n\omega})=\nonumber\\
&&\quad\frac{\sqrt[2n]{(2n-1)!!}}{n\sqrt{2\pi}\sqrt[2n]{\langle 
N_{n\omega}\rangle}N_{n\omega}^{1-\nicefrac{1}{2n}}}e^{-\frac{1}{2}\sqrt[n]{(2n-1)!!\frac{N_{n\omega}}{\langle N_{n\omega}\rangle}}}.
\label{P_nw_sb}
\end{eqnarray}
This is a heavy-tailed generalized Gamma distribution but all its moments are still finite. Eqs.~\eqref{P_nw_sb} for $n=2,3$ are plotted in Fig.~\ref{harmonics}b,c as solid black lines, after a convolution with the detector noise probability distribution (see Supplementary Information).  

The top-right insets of Fig.~\ref{harmonics}b,c show the time traces of photon numbers per pulse normalized to their mean values. We see `extreme events': a SH pulse and a TH pulse exceeding their mean values more than $200$ and $650$ times, respectively. Using the analogy of `rogue waves' in the ocean, this would correspond to a wave of about a kilometer height. This extremely `heavy-tailed' behaviour is because the number of photons in an optical harmonic pulse scales as the power function of the pump number of photons, which in our case already has a very broad distribution. In the case of supercontinuum generation, this tendency is even stronger because the dependence on the pump is exponential.      

\begin{figure*}[tb]
\begin{center}
\includegraphics[width=1\textwidth]{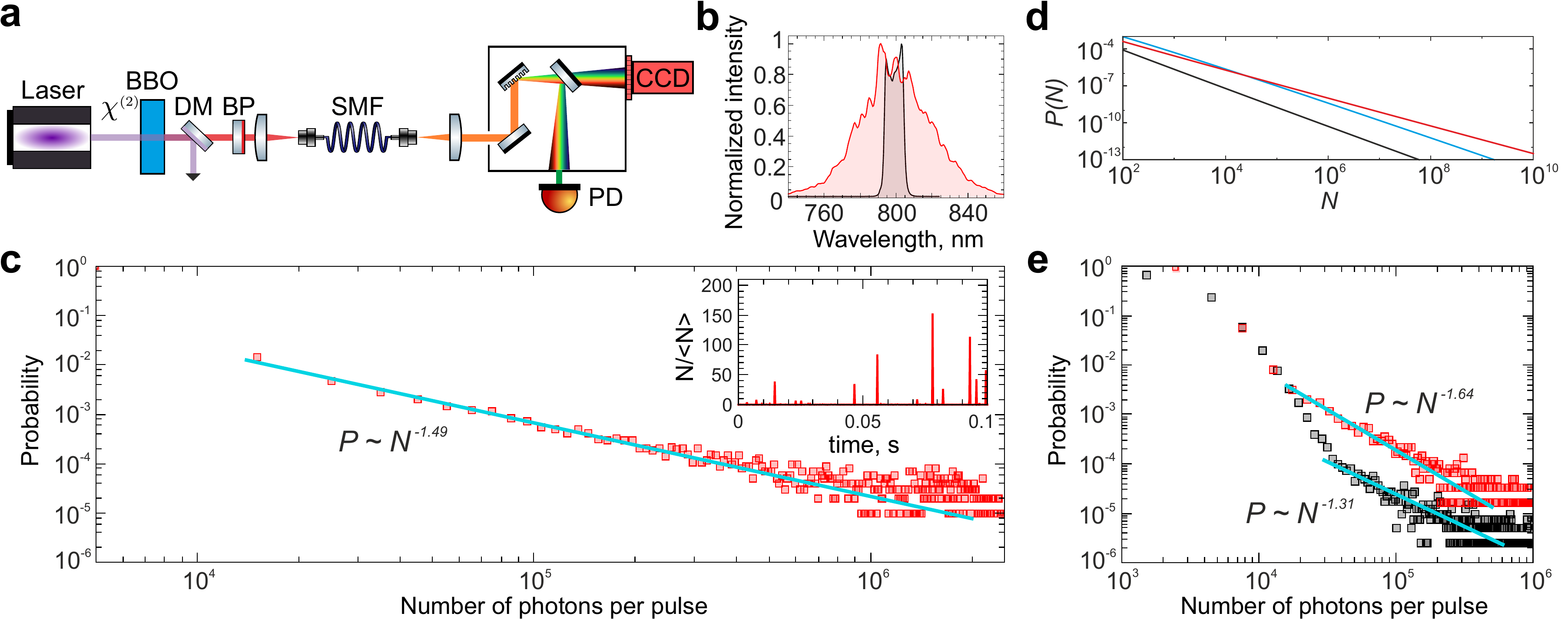}
\caption{Supercontinuum pumped by squeezed vacuum. a, the experimental setup: after BSV is generated in a BBO crystal, the pump is cut off by a dichroic mirror (DM) and BSV is filtered by a bandpass filter (BP) and coupled into a single-mode fibre (SMF); further, the supercontinuum is analyzed either by a CCD camera after a spectrometer or by a photodetector (PD) after a monochromator. b, the resulting average spectrum under pumping with $0.2$ nJ pulses (grey) and $40$ nJ pulses (pink). c, the probability distribution of photon number per pulse (points), for supercontinuum pumped with $48$ nJ pulses, and its fit with a Pareto distribution with $k=0.49$ (solid line). The inset shows a time trace for pulse height normalized to the mean. The effect of the brightness and bandwidth of BSV on the Pareto index $k$ of the supercontinuum: d, calculated photon-number distribution for single-mode BSV with $\kappa\langle N_B\rangle=0.5$ (black) and $2.5$ (red). The distribution for $\kappa\langle N_B\rangle=2.5$ and $M=5$ modes (blue) has nearly the same scaling as the one for $\kappa\langle N_B\rangle=0.5$ and $M=1$ (black). e, photon-number distributions for the supercontinuum pumped by 30~nJ pulses of BSV with bandwidth $\Delta\lambda_B=10$~nm (red points) and $\Delta\lambda_B=3$~nm (black points), and their Pareto fits (blue lines).}\label{fig:Pareto}
\end{center}
\end{figure*}
To obtain supercontinuum, we use BSV centered at $800$ nm. The initially $80$-nm broad spectrum of BSV is filtered to $10$~nm before launching it into the standard single-mode fused silica fibre, where supercontinuum is generated (Fig.~\ref{fig:Pareto}a). The spectrum after the fibre (Fig.~\ref{fig:Pareto}b) is almost unchanged (grey contour) if BSV is weak (energy per pulse $0.2$ nJ), but it is considerably broadened (pink contour) if the input energy per pulse is $40$ nJ.

The experimentally obtained probability distribution of photon number per pulse at wavelength $770$ nm under pumping with $48$ nJ pulses is displayed in Fig.~\ref{fig:Pareto}c (red points). It has a Pareto scaling with $k=0.49\pm0.02$, found from the fit of complementary cumulative distribution function (CCDF) for more accurate determination, see Supplementary Information for details. 

This shape of the probability distribution can be well explained assuming that the `blue' side of the supercontinuum is free of Raman processes and is generated through high-gain FWM. The number of photons in the supercontinuum is then an exponential function of the BSV photon number $N_B$: $N_{SC}=\sinh^2(\kappa N_{B})$, with $\kappa$ characterizing the interaction strength. 
Similar to the harmonics case, from the BSV photon-number distribution $P_B(N_{B})$ we obtain the supercontinuum one:
\begin{equation}\label{fwmsq}
P_{SC}(N)=\frac{e^{-\frac{\mathrm{arcsinh}\sqrt{N}}{2\kappa\langle N_B\rangle}}}{\sqrt{8\pi \kappa\langle N_B\rangle
N(1+N)\mathrm{arcsinh}\sqrt{N}}},
\end{equation}
where $\langle N_{B}\rangle$ is the BSV mean photon number.

For large $N$, Eq.~\eqref{fwmsq} has asymptotic scaling typical for the Pareto distribution~\cite{Foss2013}, 
\begin{equation} P(N)\propto 
\frac{1}{N^{1+k}}, \label{eq:Pareto} 
\end{equation}
with the Pareto index $k\propto M(\kappa\langle N_B\rangle)^{-1}$ (see the Supplementary Information), where $M$ is the number of modes in BSV. The Pareto index tends to zero for bright ($\langle N_B\rangle\gg1$) squeezed vacuum but increases as the number of modes $M$ grows, see Fig.~\ref{fig:Pareto}d.

To test that the Pareto index $k$ depends on both the mean photon number and the number of modes (bandwidth) of BSV, we measure the probability distributions for the supercontinuum with the BSV energy per pulse reduced to $30$ nJ. The results are shown in Fig.~\ref{fig:Pareto}e. For bandwidths $10$ and $3$ nm, the CCDF analysis gives $k=0.64\pm0.02$ and $k=0.31\pm0.02$, respectively~\footnote{The  peak below $3\cdot 10^4$ photons follows the exponential distribution; we attribute it to the thermalization effect in supercontinuum~\cite{PICOZZI2014}.}. Thus, the variation of BSV power and bandwidth are versatile instruments to control the index of the Pareto distribution for supercontinuum.

The power-law probability distributions shown in Fig.~\ref{fig:Pareto}c,e have very unusual features. With $k<1$, the mean number of photons per pulse is not defined and depends on the time of observation; it makes our distribution much different from the ones reported by the others~\cite{Borlaug2009,Kasparian2009,Alves2019}. This fractal-like behavior is typical for Pareto distributions where the mean values do not converge~\cite{Park1996}. Similar to the `coastline paradox'~\cite{Mandelbrot1967}, where the coastline appears the longer, the better one measures, the mean photon number per pulse will be the higher, the more data are used to determine it, $\langle N\rangle\propto s^{\nicefrac{1}{k}-1}$ , where $s$ is the dataset size \cite{Newman2005}. In a real experiment the distribution gets always truncated through some physical mechanisms. In our case, this occurs for photon numbers per pulse above $10^6$ due to the detector saturation.

\begin{figure}[h]
\begin{center}
\includegraphics[width=0.5\textwidth]{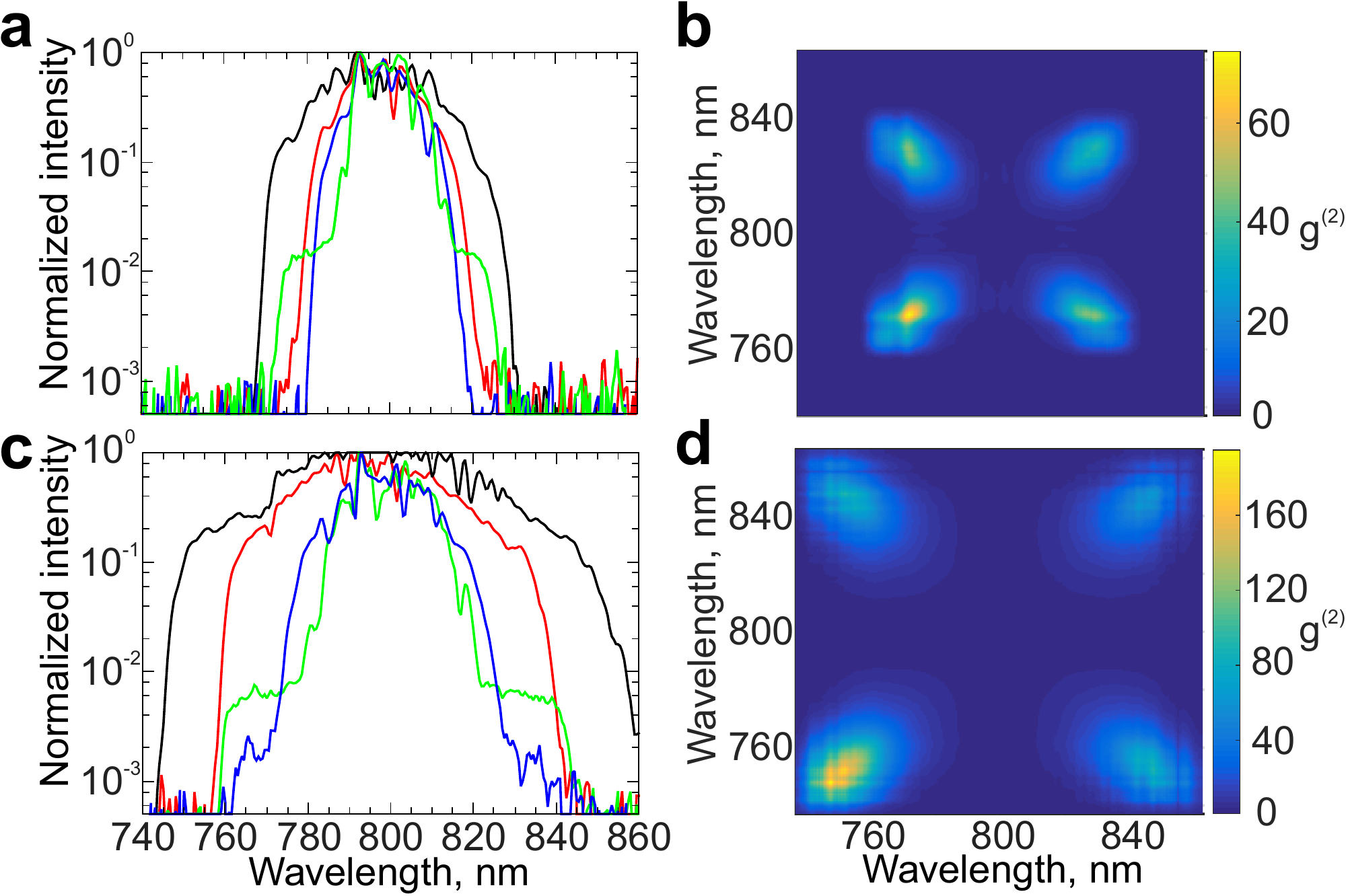}
\caption{Single-pulse spectra of supercontinuum for the input BSV energy per pulse $24$ nJ (a) and $40$ nJ (c) and the corresponding distributions of the normalized second-order correlation function (b and d, respectively).}
\label{fig:spectra}
\end{center}
\end{figure}

The power-law behaviour is only present on the supercontinuum spectra and gets suppressed in the middle (Fig.~\ref{fig:spectra}a,c). As the pulse energy increases, the spectrum broadens and the fluctuations get stronger but they also move further from the initial central wavelength. The reason is that at central frequencies, the supercontinuum is so bright that it itself generates new sidebands through FWM. But because FWM is equivalent to two-photon loss, it leads to the depletion of intensity fluctuations~\cite{Finger2017}. The same tendency is visible in the second-order normalized correlation function $g^{(2)}(\lambda,\lambda')=\langle N(\lambda)N(\lambda')\rangle/\left[\langle N(\lambda)\rangle\langle N(\lambda')\rangle\right]$ (Fig.~\ref{fig:spectra}b,d), where $N(\lambda)$ is the photon number at wavelength $\lambda$ and angle brackets denote statistical averaging. The value on the main diagonal has the meaning of the bunching parameter $g^{(2)}=\langle :N^2:\rangle/\langle N\rangle^2$, where the normal ordering can be omitted due to large $N$. In the right-bottom panel, we see $g^{(2)}$ values as high as $170$, which considerably exceeds the strongest superbunching reported to date~\cite{Spasibko2017,Meuret2017}. Similarly to the mean, the measured value depends on the time of observation and is just the lower boundary of $g^{(2)}$.

Superbunching has interesting consequences for photon subtraction experiments~\cite{Zavatta2008}, used for example 
to test Maxwell's demon in quantum thermodynamics~\cite{Vidrighin2016}. In a photon subtraction experiment, a quantum state of light $|\psi\rangle$ is fed to a beamsplitter, after which a single-photon detector can register a reflected photon. Provided that the detector registers a single photon, the state of light after the beamsplitter is photon-subtracted, $|\psi'\rangle\propto\boldsymbol{\hat{a}}|\psi\rangle$, where $\boldsymbol{\hat{a}}$ is the 
photon annihilation operator. Counterintuitively, the photon-subtracted state has the mean photon number increased by a factor $g^{(2)}$ compared to the initial state~\cite{Bogdanov2017,Hlouvsek2017}. This energy enhancement admits an interpretation similar to Maxwell's demon experiment~\cite{Vidrighin2016}. 

\begin{figure}[h]
\begin{center}
\includegraphics[width=0.5\textwidth]{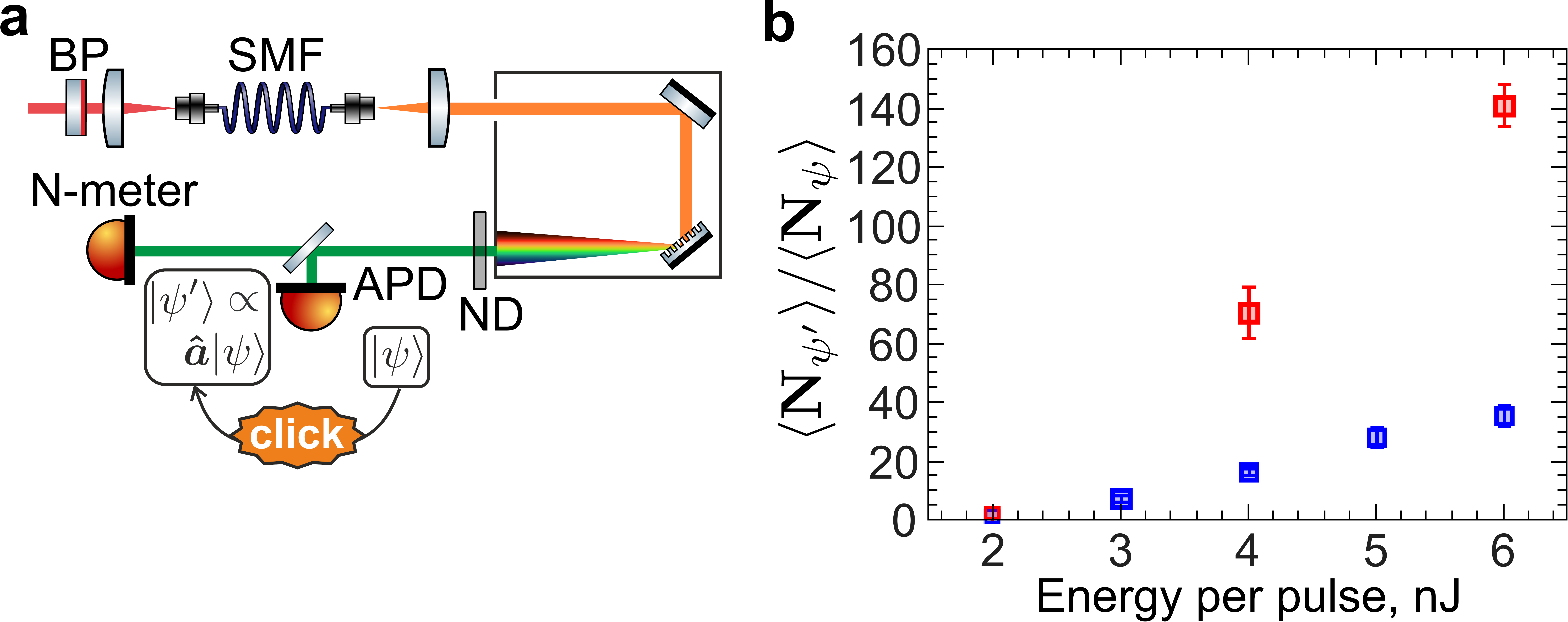}
\caption{Photon subtraction from supercontinuum. a, the experimental setup:  BSV is filtered from the pump by a bandpass filter (BP) and 
sent to the fibre. At the output, the supercontinuum is filtered by a monochromator, attenuated by a neutral-density filter (ND) and sent to a 
beamsplitter.  Only if the avalanche photodiode (APD) registers a photon, the number of photons is measured at the output (N-meter). b, the increase of the output mean photon number $\langle N_{\psi'}\rangle/\langle N_{\psi}\rangle$ after the subtraction of a single photon ($|\psi'\rangle\propto\boldsymbol{\hat{a}}|\psi\rangle$),  as a function of the mean energy per pulse of the input BSV for $10$ nm (blue) and $3$ nm (red) bandwidth.}\label{subtraction} \end{center}
\end{figure}

Our  photon  subtraction setup is shown in Fig.~\ref{subtraction}a. After the spectral filtering of the supercontinuum down to a $1$~nm bandwidth at wavelength 780~nm, the output of the monochromator is attenuated with a neutral density filter in order to  obtain on average much less than one photon ($\langle N_{\psi}\rangle\ll1$) before the beamsplitter. The mean number of photons $\langle N_{\psi'}\rangle$ after the beamsplitter is measured with an avalanche photodiode (an $N$-meter) under the condition that a single photon is detected in the reflected arm by another avalanche photodiode (APD). As a result of photon subtraction, the mean photon number is increased up to 140 times (Fig.~\ref{subtraction}b), depending on the bandwidth and the energy per pulse of the BSV pumping the supercontinuum generation. This drastic increase in the mean photon number shows that the supercontinuum can be brought out of equilibrium by the subtraction process much more efficiently than thermal or BSV light. As a result, much more work could be, in principle, extracted, which makes the supercontinuum a useful resource for proof-of-principle tests of quantum thermodynamics \cite{Hlouvsek2017}.

This radiation with strong fluctuations in the photon number can be also useful in ghost imaging~\cite{Pittman1995,Gatti2004} where the contrast of the image is given by the bunching parameter $g^{(2)}$~\cite{Agafonov2009,Chan2009}. In particular, recently time-domain ghost imaging has been demonstrated with incoherent supercontinuum, although without high photon-number fluctuations~\cite{Amiot2018}. The use of supercontinuum with Pareto photon distribution will drastically increase the contrast. 

In conclusion, we have demonstrated heavy-tailed distributions of photon numbers for nonlinear effects generated from BSV. In particular, supercontinuum generation leads to an extremal Pareto photon-number distribution with indefinite first and higher moments. The mechanism behind this phenomenon is the exponential dependence of the frequency conversion efficiency on the photon number of the pump, which has already a very broad distribution. Depending on the mean photon number and bandwidth of BSV, we obtain the Pareto index equal to $0.31$, $0.49$, and $0.64$. All these values lead to a photon-number (pulse-energy) probability distributions with diverging mean value, which has been never observed before. The bunching parameter we observe in this case is as high as $170$ and far exceeds the ones reported earlier in the literature. The subtraction of a single photon from the supercontinuum increases the mean number of photons by more than two orders of magnitude, which can be further exploited to extract large amount of thermodynamical work.

We thank Felix Thoma and Denis Kopylov for helping with the experiment at early stages. R.F. thanks \' Eva R\' acz and L\' aszlo Ruppert for helping with the tail exponent methods. M.V.C. thanks Nail Akhmediev for helpful discussions. K.Yu.S. thanks Vera Beletckaia for helpful discussions. We acknowledge the financial support of the joint DFG--RFBR (Deutsche Forschungsgemeinschaft -- Russian Foundation for Basic Research) Project No. CH1591/2-1 -- 16-52-12031 NNIOa and of DAAD (Deutsche Akademische Austauschdienst) PPP Tschechien 2017 Project No. 57319488. R.F. acknowledges project GB14-36681G of the Czech Science Foundation and Bilateral research grant 7AMB17DE034 of the Czech Ministry of Education and German Academic Exchange Service (DAAD).


%

\setcounter{figure}{0}
\renewcommand\thefigure{S\arabic{figure}}
\renewcommand\thetable{S\arabic{table}}

\section{Supplementary information to ``Indefinite-mean Pareto photon distribution from amplified quantum noise''}
\subsection{Experimental details for the generation of BSV, optical harmonics, and supercontinuum}
\setcounter{page}{1}

{\it Generation of BSV.} For the experiments with optical harmonics, BSV is generated in a $10$ mm beta barium borate (BBO) crystal through type-I collinear frequency-degenerate PDC~\cite{Spasibko2016} pumped by 1.6 ps pulses of regeneratively amplified Ti-sapphire laser at $800$ nm with a $5$ kHz repetition rate and energy per pulse up to $0.5$ mJ (Fig.~\ref{harmonics}a). To reduce the effect of spatial walk-off~\cite{Perez2015}, the pump beam is focused into the crystal with a cylindrical lens. The resulting BSV spectrum is centered at $1600$ nm (Fig.~\ref{BSV}a). For the supercontinuum generation, BSV at $800$ nm is used, generated through type-I collinear degenerate PDC in two 3 mm BBO crystals~\cite{Manceau2017} from the frequency doubled radiation of the same laser (wavelength $400$ nm, energy per pulse up to 0.2 mJ). After cutting off the pump with a dichroic mirror, BSV is filtered spatially and spectrally: with a slit and 4f monochromator down to a single mode (Fig.~\ref{harmonics}a) or with a bandpass filter (Fig.~\ref{fig:Pareto}a) to a few-mode case, respectively. In the latter case the fibre itself provides single-mode spatial filtering.

{\it Generation of optical harmonics.}  Second harmonic (SH) and third harmonic (TH) are generated by tightly focusing BSV on the surface of a 1 mm slab of lithium niobate crystal, with the $z$ crystal axis in the plane of the slab and the BSV polarized along $z$. Under this condition, the largest components of second and third nonlinear susceptibilities are used, providing a high efficiency even though phase matching is not satisfied~\cite{Spasibko2017,Kopylov2019} (Fig.~\ref{harmonics}a). The photon number distribution of BSV is measured by a charge-integrating detector based on an infrared p-i-n diode, providing a linear response up to $10^6$ photons per pulse and noise equivalent to $1600$ photons per pulse. The photon number distribution for the harmonics is measured with a similar visible detector, with the noise equivalent to $270$ photons per pulse~\cite{Iskhakov2009}. After cutting off BSV with a short-pass filter, each harmonic is additionally filtered with bandpass filters at $800$ nm (SH) or $532$ nm (TH) to block the other one. 

{\it Supercontinuum is generated} in a $5$ m single-mode fused silica  fibre and further, to achieve single-mode detection, spectrally filtered with a monochromator whose resolution is $1$ nm (Fig.~\ref{fig:Pareto}a). The photon-number probability distribution is measured by a visible charge-integrating detector (the same as in the case of optical harmonics). The single-pulse spectra (Fig.~\ref{fig:spectra}a,c) are recorded by a CCD camera at the output of a spectrometer and further processed to measure the second-order normalized correlation function $g^{(2)}(\lambda,\lambda')$ shown in Fig.~\ref{fig:spectra}b,d. 

{\it Taking detection noise into account.}
The dark noise of charge-integrating photodetectors is well described by a Gaussian distribution with zero mean,
\begin{equation}
P(N) = \frac{1}{\sigma\sqrt{2\pi}}e^{-\frac{N^2}{2\sigma^2}},
\label{P_gauss}
\end{equation} 
with the standard deviation $\sigma$. Because this noise is independent from the fluctuations of the detected light, it 
just adds to the photon-number noise.

The total probability distribution is given by the convolution of the photon-number and dark noise probability distributions. The convolution of Eq.~\eqref{P_gauss} with Eq.~\eqref{P_nw_sb}, with no fitting parameters, perfectly coincides with the experimental histograms for the generated harmonics (Fig.~\ref{harmonics}b,c). 

\subsection {Analysis of the complementary cumulative distribution functions} 
The tails of probability distribution functions (PDFs) are usually analyzed through their complementary cumulative distribution functions 
(CCDFs)~\cite{Beirlant2005},
\begin{equation}
\bar{C}(N)=\int_{N}^{\infty} P(N')dN',
\label{CCDF}
\end{equation}
whose analysis is possible even if the moments diverge.

The tail index of a PDF is defined as 
\begin{equation}
\alpha=\lim_{N\rightarrow\infty} \frac{H(N)}{N}, 
\label{alpha}
\end{equation}
where $H(N)=-\log\bar{C}(N)$ is called the hazard function. If $\alpha=\mathrm{const}$, the distribution decays exponentially. For diverging 
$\alpha$ the tail decays faster than exponential, for $\alpha=0$ slower. The latter means a heavy-tailed distribution~\cite{Foss2013}.

\begin{figure}[h]
\begin{center}
\includegraphics[width=0.45\textwidth]{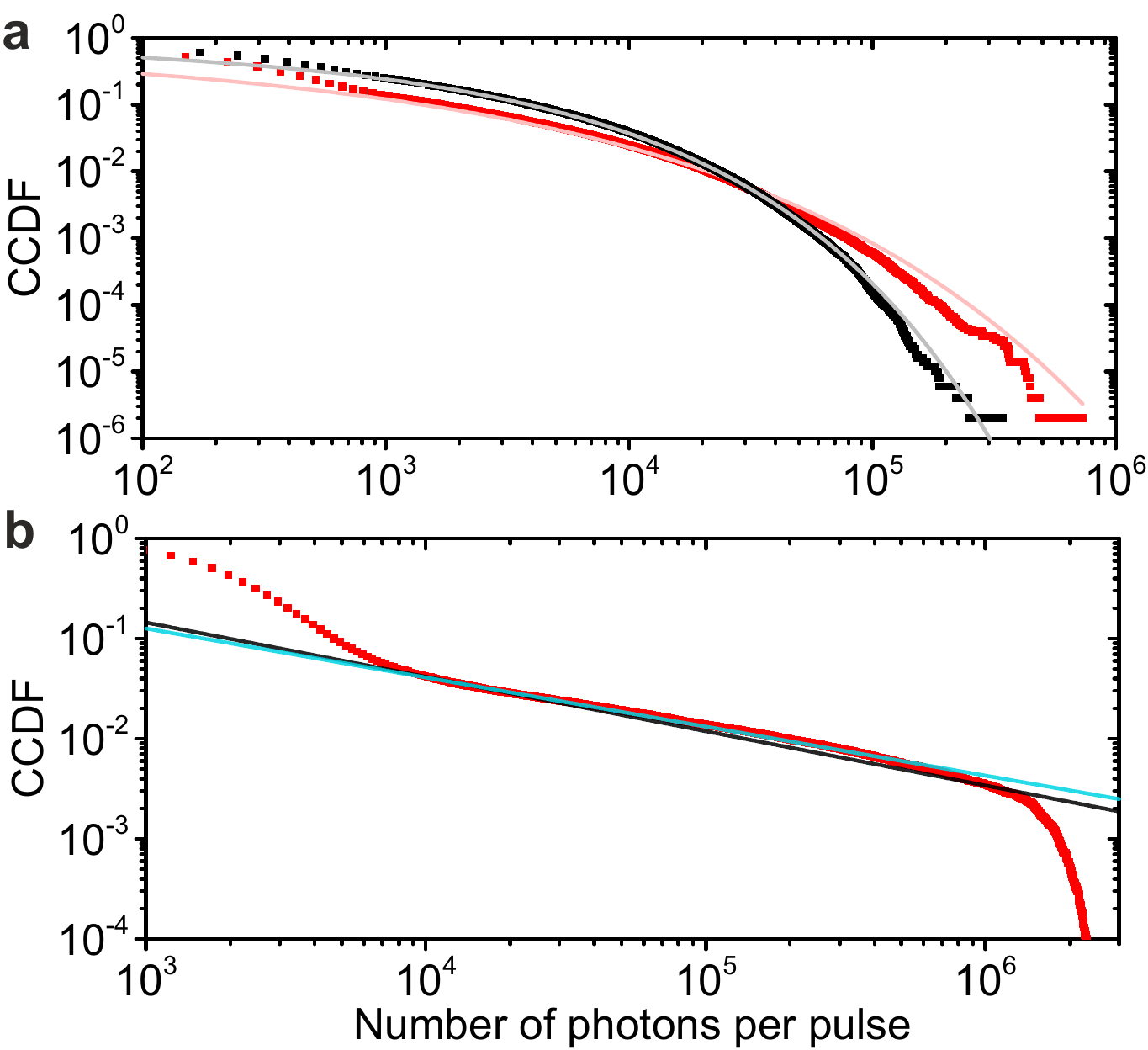}
\caption{(a) CCDFs for the data on optical harmonics from Fig.~\ref{harmonics}b,c: $2\omega$ (black points) and $3\omega$ (red points). The corresponding theoretical distributions [\eqref{CCDF_nw_sb} for $n=2,3$] are shown by gray and pink lines respectively.  $\langle N_{n\omega}\rangle$ are taken from the experimental data. (b) CCDF for the data on the supercontinuum from Fig.~\ref{fig:Pareto}c (red points), its fit (blue line) with Eq.~\eqref{eq:Pareto} leading to $k_e=0.49$, and Eq.~\eqref{eq:Pareto} with $k=0.53$ (black line) obtained from the maximum likelihood estimator~\cite{Newman2005,Clauset2009}.}
\label{fig:CCDF}
\end{center}
\end{figure}

\begin{figure}[h]
\begin{center}
\includegraphics[width=0.4\textwidth]{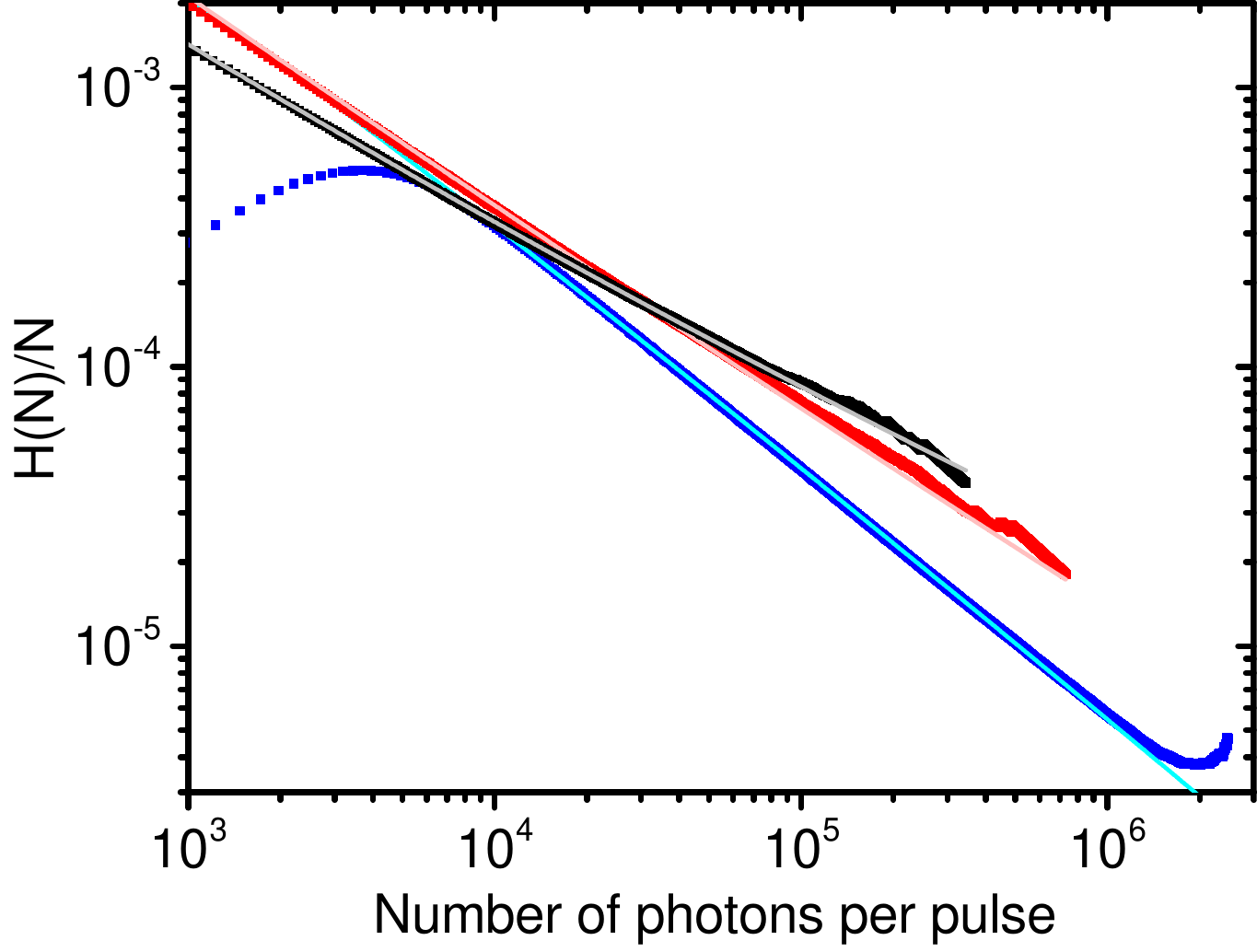}
\caption{Experimental (points) and theoretical (lines) $H(N)/N$ for second (black) and third (red) harmonics from Fig.~\ref{harmonics}b,c and supercontinuum (blue) from Fig.~\ref{fig:Pareto}c. For the latter, the theoretical values are calculated from a fit with the Pareto distribution (Fig.~\ref{fig:CCDF}b).}
\label{fig:TailIndexAlpha}
\end{center}
\end{figure}

For the $n$th optical harmonic of BSV, with the PDF given by (\ref{P_nw_sb}), the CCDF is
\begin{equation}
\bar{C}_{n\omega}(N_{n\omega})=\mbox{Erfc}\left[\frac{\sqrt[2n]{(2n-1)!!\frac{N_{n\omega}}{\langle N_{n\omega}\rangle}}}{\sqrt{2}}\right],
\label{CCDF_nw_sb}
\end{equation}
where $\mbox{Erfc}(x)$ is the complementary error function. CCDFs for SH and TH, together with the experimental data, are presented in Fig.~\ref{fig:CCDF}a. From these dependences and the corresponding $H(N)/N$ dependences (Fig.~\ref{fig:TailIndexAlpha}), one can see that $\alpha$ tends fast to zero, therefore the distributions are heavy-tailed.

In the case of supercontinuum generated from BSV, the PDF is given by \eqref{fwmsq}, and the corresponding CCDF is
\begin{equation}
\bar{C}_{SC}(N)=\mbox{Erfc}\left[\sqrt{\frac{\mathrm{arcsinh}\sqrt{N}}{2\kappa\langle N_B\rangle}}\right].
\label{CCDF_scg_sb}
\end{equation}
Similarly to the case of harmonics, $\alpha$ tends to zero, the distribution is heavy-tailed. However, it exhibits a faster tendency to zero than for any harmonic (Fig.~\ref{fig:TailIndexAlpha}).
The CCDF \eqref{CCDF_scg_sb} is of the form $N^{-k}L(N)$, where $k$ is the tail exponent and $L(N)$ is a  slowly varying 
function (i.e.~$\lim_{N\rightarrow \infty} \nicefrac{L(tN)}{L(N)}=1$, for any $t>1$ \cite{Foss2013}). Therefore distribution \eqref{fwmsq} belongs to regularly varying distributions with a finite tail exponent. It is tail equivalent to the Pareto distribution with the same $k$. The tail exponent tends to $(4\kappa\langle N_B\rangle)^{-1}$. For a sufficiently strong pump, $k$ goes below unity and then, all moments are indefinite. It makes \eqref{fwmsq} very different from the other heavy-tailed distributions.

The tail exponents linearly increase with the number of modes. Indeed, BSV with $M$ modes has
\begin{equation}\label{eq:P_sb_M}
P_{B,M}(N_B)=\frac{N_B^{\nicefrac{M}{2}-1}}{\Gamma\left(\nicefrac{M}{2}\right)}\left(\frac{M}{2\langle N_B\rangle}\right)^{\nicefrac{M}{2}}e^{-\frac{MN_B}{2\langle N_B\rangle}},
\end{equation}
where $\Gamma(x)$ is the gamma function. The corresponding CCDF for supercontinuum is
\begin{equation}\label{eq:C_scg_sb_M}
\bar{C}_{SC,M}(N)=\frac{\Gamma\left(\frac{M}{2},\frac{\mathrm{arcsinh}\sqrt{N}}{2\kappa\langle N_B\rangle/M}\right)}{\Gamma\left(\nicefrac{M}{2}\right)},
\end{equation}
where $\Gamma(s,x)$ is the upper incomplete Gamma function. The latter leads to
\begin{equation}\label{eq:k_t}
k=\frac{M}{4\kappa\langle N_B\rangle}.
\end{equation}

Remarkably, the distribution~\eqref{eq:P_sb_M} for $M=2$ becomes exactly the exponential one. As a result, Eq.~\eqref{eq:k_t} immediately tells us that with a BSV pump, $k<1$ is achieved with twice lower mean photon number than with a thermal pump.

Experimental tail exponents $k_e$ are derived from the fits of the linear part of CCDFs in the log-log scale. The error of the fit $\Delta k_{e}=0.02$ is mainly caused by imprecise estimation of the starting and ending points of this range. The other method, based on the calculation of the maximum likelihood estimator~\cite{Newman2005,Clauset2009}, provides similar exponents in our case. An example of CCDF with both 
fits and the resulting exponents are presented in Supplementary Figure~\ref{fig:CCDF} and Table~\ref{Table_SCG}.

\begin{table}[h]
\renewcommand{\arraystretch}{1.25}
\begin{tabular}{|c|c|c|c|c|}
\hline
& $P$, $\Delta\lambda_B$&$k_{e}$&$k_{t}$&$\langle N\rangle$, photons/pulse\\
\hline
Fig.~\ref{fig:Pareto}c&48~nJ, 10 nm&$0.49$&$0.31$&$1.18\times10^4$\\
\hline
Fig.~\ref{fig:Pareto}e&30~nJ, 10 nm&$0.64$&$0.5$&$6.5\times10^3$\\
\cline{2-5}
&30~nJ, 3 nm&$0.31$&$0.2$&$5.6\times10^3$\\
\hline
\end{tabular}
\caption{Characteristics of the supercontinuum photon-number distributions from Figs.~\ref{fig:Pareto}c,e. $P$ and $\Delta\lambda_B$ are 
BSV energy per pulse and bandwidth, respectively; $k_{e}$ and $k_{t}$ are experimental and theoretical tail exponents. Despite $k$ being less than 1, the mean number of photons $\langle N\rangle$ exists for the measured distributions due to truncation.} \label{Table_SCG}
\renewcommand{\arraystretch}{1}
\end{table}

The theoretical exponents $k_{t}$, estimated from Eq.~\eqref{eq:k_t}, are somewhat smaller than the experimental ones. We get the mean gain value ($\kappa\langle N_B\rangle=4\pm0.5$ and $2.5\pm0.5$ for $P=48$ and 30~nJ, respectively) from the nonlinear dependence, similar to the one for BSV (Fig.~\ref{BSV}b). The number $M$ of modes ($M=5\pm1$ and $2\pm0.2$ for $\Delta\lambda_B=10$ and 3~nm, respectively) is obtained from the $g^{(2)}$ value \cite{Iskhakov2012} for extremely weak BSV ($P=3$~pJ) after the fibre. The uncertainties in both measurements result in the relative error $\delta k_{t}=25\%$.

\clearpage

\end{document}